\documentclass[12pt]{article}
\pdfoutput=1
\usepackage{putex}
\usepackage{feyn}
\usepackage[vcentermath]{youngtab}

\usepackage{graphicx}
\usepackage{epstopdf}
\usepackage{enumerate}
\usepackage{cite}
\usepackage{tensor}
\usepackage{slashed}
\usepackage{feynmf}

\usepackage{hyperref}

\numberwithin{equation}{section}

\newcommand{\HH}{\mathbb{H}}

\newcommand {\be} {\begin {equation}}
\newcommand {\ee} {\end {equation}}

\newcommand {\bes} {\begin {equation*}}
\newcommand {\ees} {\end {equation*}}

\newcommand{\es}[2] {\begin{equation} \label{#1} \begin{split} #2 \end{split} \end{equation}}

\def\CO{{\cal O}}

\newcommand{\beq}{\begin{equation}}
\newcommand{\eeq}{\end{equation}}

\def\be{ \begin{equation} }
\def\ee{ \end{equation} }

\begin{document}

\preprint{PUPT-2451}

\institution{PU}{Department of Physics, Princeton University, Princeton, NJ 08544}
\institution{PCTS}{Princeton Center for Theoretical Science, Princeton University, Princeton, NJ 08544}

\title{
One Loop Tests of Higher Spin AdS/CFT
}

\authors{Simone Giombi\worksat{\PU} and Igor R.~Klebanov\worksat{\PU,\PCTS}}

\abstract{
Vasiliev's type A higher spin theories in $AdS_4$ have been conjectured to be dual to the $U(N)$ or $O(N)$ singlet sectors in 3-d conformal
field theories with $N$-component scalar fields.
We compare the ${\cal O}(N^0)$ correction to the 3-sphere free energy $F$ in the CFTs
with corresponding calculations in the higher spin theories. This requires evaluating a regularized sum over one loop vacuum energies of
an infinite set of massless higher spin gauge fields in Euclidean $AdS_4$. For the Vasiliev theory including fields of all integer spin
and a scalar with $\Delta=1$ boundary condition, we show that the regularized
sum vanishes. This is in perfect agreement with the vanishing of subleading corrections to $F$ in the $U(N)$ singlet sector of the theory of $N$ free complex scalar fields. For the minimal Vasiliev theory including fields of only even spin,
the regularized sum remarkably equals the value of $F$ for one free real scalar field.
This result may agree with the
$O(N)$ singlet sector of the theory of $N$ real scalar fields, provided the coupling constant in the Vasiliev theory is identified as $G_N\sim 1/(N-1)$.
Similarly, consideration of the $USp(N)$ singlet sector for $N$ complex scalar fields, which we conjecture to be dual to the
$husp(2;0|4)$ Vasiliev theory, requires $G_N\sim 1/(N+1)$.
We also test the higher spin $AdS_3/CFT_2$ conjectures by calculating the regularized sum over one loop vacuum energies of higher spin fields in $AdS_3$. We match the result with the ${\cal O}(N^0)$ term in the central charge of
the $W_N$ minimal models; this requires a certain truncation of the CFT operator spectrum so that the bulk theory contains two real scalar fields
with the
same boundary conditions.
}

\date{}
\maketitle

\tableofcontents

\section{Introduction and Summary}

The existence of massless interacting theories of higher spin gauge fields in AdS space \cite{Fradkin:1987ks,Vasiliev:1990en,Vasiliev:1992av,Vasiliev:1995dn,Vasiliev:1999ba}
is a fascinating fact that has stimulated a lot of recent research.
In particular, it allows for extensions of the AdS/CFT correspondence \cite{Maldacena:1997re,Gubser:1998bc,Witten:1998qj}
to non-supersymmetric large $N$ field theories where the dynamical fields are $N$-vectors rather than $N\times N$ matrices \cite{Klebanov:2002ja}.
The $U(N)$ singlet sector of the 3-d theory of $N$ massless complex scalar fields has been conjectured to be dual to Vasiliev's type A theory in
$AdS_4$, which contains each integer spin once \cite{Klebanov:2002ja}.
The latter theory has a truncation to a so-called minimal theory containing even spins only; it has been conjectured to be dual to the the $O(N)$
singlet sector of the 3-d theory of $N$ massless real scalar fields. Similarly, 3-d theories of $N$ massless fermion fields have been conjectured to be
dual to Vasiliev's type B theory, where the scalar has negative parity \cite{Sezgin:2003pt,Leigh:2003gk}. Comparisons of
three-point correlation functions have provided considerable evidence in favor of these conjectures \cite{Giombi:2009wh,Giombi:2010vg,Giombi:2011ya,Maldacena:2011jn,Maldacena:2012sf,Colombo:2012jx,Didenko:2012tv,Gelfond:2013xt,Didenko:2013bj}. Another important recent development \cite{Aharony:2011jz,Giombi:2011kc} has been the realization that
the $U(N)$ or $O(N)$ singlet constraint may be imposed by coupling the dynamical massless fields to a Chern-Simons theory at level $k$.
When $k$ is sent to infinity, then the duality conjectures for Vasiliev's type A and B models are recovered. When the 't~Hooft coupling $N/k$ is kept finite, the bulk duals are conjectured \cite{Giombi:2011kc, Chang:2012kt} to be more general parity breaking Vasiliev's theories, whose equations of motion contain extra parameters which are mapped to the CFT 't~Hooft coupling.
Calculation of correlation functions at leading order in $N$ provides evidence for these
generalized vectorial AdS/CFT dualities \cite{Giombi:2012ms}.

The goal of our paper is to test the above conjectures in a new way by considering the ${\cal O}(N^0)$ correction to the CFT partition function.
By the AdS/CFT dictionary, the CFT partition function $F=-\log Z$ on a round $S^3$ is related to the partition function of the bulk theory on the Euclidean $AdS_4$ vacuum with metric
\begin{equation}
ds^2=d\rho^2+\sinh^2\rho\, d\Omega_3\,,
\label{AdS4}
\end{equation}
where $d\Omega_3$ is the metric of a unit round 3-sphere. An interesting challenge is to try to match the $F$ value for $N$
free conformal 3-d fields with that in the Vasiliev higher spin theory in the Euclidean $AdS_4$ \cite{Klebanov:2011gs}.
On general grounds, the bulk partition function takes the form
\begin{equation}
Z_{\rm bulk} = e^{-\frac{1}{G_N} F^{(0)}-F^{(1)}-G_N F^{(2)}+\ldots} = e^{-F_{\rm bulk}}\,,
\label{Zbulk}
\end{equation}
where $G_N$ denotes the bulk coupling constant. By a simple argument based on matching the large $N$ scaling of correlation functions, one sees that $G_N^{-1} \propto N$ in the large $N$ limit.\footnote{This is because the dynamical fields in the dual CFT are $N$-vector fields rather than $N\times N$ matrices.}

The leading term in (\ref{Zbulk}) corresponds to the value of the suitably regularized classical action computed on the vacuum solution with metric
(\ref{AdS4}), and with all other fields set to zero. Since we do not currently know an action for Vasiliev's theory which reduces to the
standard quadratic actions at the free level,\footnote{Actions for Vasiliev's theory were proposed in
\cite{Boulanger:2011dd,Boulanger:2012bj,Doroud:2011xs} (see \cite{Vasiliev:1988sa} for earlier work). They do not appear to reduce to ordinary quadratic lagrangians when expanded around the $AdS_4$ vacuum, and it is not clear to us how to extract the tree level free energy $F^{(0)}$ from those actions.} a direct calculation of $F^{(0)}$ appears to be hard and we will not address it here. The subleading terms in (\ref{Zbulk}) arise from the quantum corrections, and are obtained by computing the vacuum diagrams of the bulk fields in the $AdS_4$ background. In particular the calculation of the one loop term $F^{(1)}$, which is the focus of this paper, is in principle a well-posed problem.\footnote{This problem is harder, though, than calculating the difference between the
values of $F^{(1)}$ corresponding to the two possible boundary conditions in AdS space. These calculations were carried out in
\cite{Gubser:2002zh,Gubser:2002vv,Hartman:2006dy,Diaz:2007an,Klebanov:2011gs,Giombi:2013yva}.}
 We have to evaluate the one loop determinants for all bulk fields propagating in the $AdS_4$ background, and then
regularize their sum so that the power law divergences are removed. In practice, we perform this regularization using a certain analytic continuation
similar to the $\zeta$ function regularization.

The results, which we present in section 2, turn out to be quite interesting. For the theory of $N$ free complex fields, $F$ is of order $N$ and all the higher order corrections
should vanish. We show that the ${\cal O}(N^0)$ correction $F^{(1)}$ indeed vanishes in the non-minimal type A Vasiliev theory containing fields of all integer spin
including a scalar with the $\Delta=1$ boundary condition.
Calculation in the minimal type A theory containing fields of even spin only, gives instead
$F^{(1)}_{{\rm min}}=\frac{\log 2}{8}-\frac{3\zeta(3)}{16\pi^2}$. Remarkably, this is exactly the same as
the contribution to $F$ of a single real conformal scalar field \cite{Klebanov:2011gs}!\ On the other hand, in the theory of $N$ real free fields, the ${\cal O}(N^0)$ correction
to $F$ of course must vanish. We believe that the bulk calculation can be consistent with this due to a subtlety in the definition of
$G_N$. The exact relation between $G_N$ an $N$ may involve corrections which are subleading at large $N$. It was argued in \cite{Maldacena:2011jn} that the coupling constant in Vasiliev's theory should be quantized. The most general relation between $G_N$ and $N$ should then be of the form $G_N^{-1}=\gamma (N+n)$, where $\gamma$ is a constant and $n$ a fixed integer. For the duality involving the minimal type A theory and $O(N)$ invariant theory of real scalar fields, taking $n=-1$ can
make the bulk calculation fully consistent with field theory expectations. In fact, exactly the same shift was found for the topological string theory dual to
$SO(N)$ Chern-Simons gauge theory \cite{Sinha:2000ap}.

In section 2.1 we conjecture a new higher-spin duality which relates the $USp (N)$ singlet sector of the theory of $N$ free complex scalar
fields (here $N$ is even) to the $husp(2;0|4)$ Vasiliev theory in $AdS_4$. The latter theory contains one field of each even spin and three
fields of each odd spin \cite{Vasiliev:1999ba}. Using this spectrum we find that  $F^{(1)}_{husp(2;0|4)} =-\frac{\log 2}{4}+\frac{3\zeta(3)}{8\pi^2}$, i.e. minus the 3-sphere free energy of a single complex scalar field. This means that taking $G_N^{-1}=\gamma (N+1)$ for this theory can provide consistency with the conjectured AdS/CFT duality.
The shift $N\rightarrow N+1$ was also found for the topological string theory dual to
$USp(N)$ Chern-Simons gauge theory \cite{Sinha:2000ap}.

One may be concerned that our AdS/CFT matching of $F$ is incomplete due to the presence of the $U(N)$ or $O(N)$ or $USp(N)$ Chern-Simons gauge fields.
They give contributions to $F$ which are of order $N^2 \log (k/N)$ \cite{Klebanov:2011gs}, and there are no terms of this order
in the Vasiliev theory.\footnote{The physical effects of Chern-Simons gauge fields on manifolds of higher topology were studied in
\cite{Banerjee:2012gh}.}
Indeed, in the complete formulation of the bulk dual of the Chern-Simons matter theories, the Vasiliev theory probably has to be coupled to
a topological sector that is the dual of the pure Chern-Simons theory\cite{JHV,Giombi:2012ms}. Such topological string theories have been constructed in
\cite{Gopakumar:1998ki,Sinha:2000ap}. While the coupling of Vasiliev theory to the topological sector is not clear to us, we believe that
it does not affect our matching in the limit $k\rightarrow \infty$. In this limit, the contribution to $F$ from $N$ complex scalars has the structure
$N \left (\frac{\log 2}{4}-\frac{3\zeta(3)}{8\pi^2}\right )(1+ {\cal O}(N/k))$, so that the corrections due to Chern-Simons interactions become negligible. More precisely, the free energy receives contributions from two types of diagrams: those that involve only gauge field propagators, and those that involve at least one matter propagator. The former correspond to the pure Chern-Simons theory, and their contribution diverge as $N/k\rightarrow 0$. The latter arise from Chern-Simons matter interactions, and they have a smooth limit as $N/k\rightarrow 0$. One may hope that, once the pure Chern-Simons contribution is subtracted, the bulk Vasiliev theory correctly captures the contribution of the second class of diagrams, at least in the $k\rightarrow \infty$ limit.

However, for non-vanishing $N/k$ the Chern-Simons coupling becomes important, and it may be harder to make a sharp comparison. The conjectures of \cite{Giombi:2011kc, Chang:2012kt} relate the Chern-Simons matter theories at finite 't~Hooft coupling to certain parity violating higher spin theories. However, the free spectrum of these theories is unchanged when turning on the parity breaking interactions,\footnote{The parity breaking Vasiliev theories depend on certain phase-like parameters which are functions of the `t~Hooft coupling $\lambda=N/k$ and affect higher spin interactions, but not the free equations of motion. However, in these theories the bulk coupling constant $G_N$ is also expected to be a function of the 't~Hooft coupling of the form $G_N^{-1} \sim N \frac{\sin\pi\lambda}{\pi \lambda}$, in order to match the $\lambda$ dependence of the 2-point function of the stress tensor (and higher spin currents) in the CFT \cite{Aharony:2012nh,GurAri:2012is}.} which would lead to the same result for the bulk one loop free energy as in the parity preserving Vasiliev's theory discussed above. On the other hand, the ${\cal O}(N^0)$ contribution to the CFT free energy is expected to be a non-trivial function of the coupling $N/k$.  A similar puzzle is posed by the type B theories
which are supposed to be dual to Chern-Simons theory coupled to one fermion in the fundamental representation.
In this case the fermion provides a half-integer shift of $k$ due to the parity anomaly \cite{Niemi:1983rq,Redlich:1983kn}, so
it may be inconsistent to simply subtract the Chern-Simons contribution. Indeed, the comparison with the one loop correction in Vasiliev theory, that works so well for
the singlet sector of scalar theories, does not yield agreement for the fermionic theories.
For the bulk theory containing all integer spin fields and a scalar with the $\Delta=2$ boundary condition, we find
$F^{(1)}=- {\zeta(3)\over 8 \pi^2}$. This value is correct for the critical $O(N)$ model \cite{Klebanov:2011gs}, but it does not agree with
the vanishing result expected in the theory of $N$ Dirac fermions.
We hope to return to these puzzles in the future.

In section 3 we carry out a one loop test of higher-spin $AdS_3/CFT_2$ dualities. A conjecture by Gaberdiel and Gopakumar \cite{Gaberdiel:2010pz,Gaberdiel:2012uj}
relates the 2-d
$\frac{SU(N)_k\otimes SU(N)_1}{SU(N)_{k+1}}$ coset CFTs, known as the $W_N$ minimal models, to higher-spin theory in $AdS_3$ \cite{Prokushkin:1998bq,Vasiliev:1999ba}.
Working in the large $N$ `t Hooft limit
we reproduce the ${\cal O}(N^0)$ correction to the
central charge of this theory
using a one loop calculation in $AdS_3$. In doing so, we have to assume that the bulk theory contains
gauge fields of spin $s=2, 3, \ldots$, as well as two real scalar fields corresponding to CFT operators of the same dimension $\Delta_\pm= 1\pm \lambda$.
This is exactly the spectrum that appears in certain truncations of the coset CFT that were proposed by Chang and Yin \cite{Chang:2011mz}.
If the `t Hooft coupling is identified as $\lambda=\frac{N}{N+k}$ then we have to adopt the
$\Delta_-$ boundary conditions; if it is instead identified as $\lambda=\frac{N}{N+k+1}$ \footnote{We are grateful to Rajesh
Gopakumar for informing us about this possibility.} then we have to adopt
$\Delta_+$.
A refinement of the original conjecture of \cite{Gaberdiel:2010pz} was also presented in \cite{Gaberdiel:2012ku}, with a similar conclusion that the perturbative spectrum of the bulk theory should only include two real
scalar fields with the same choice of boundary condition, i.e. one complex field as in the bosonic truncation of the models of \cite{Prokushkin:1998bq,Vasiliev:1999ba}.
The analysis of \cite{Gaberdiel:2012ku} favors the $\Delta_{+}$ boundary condition for the two perturbative bulk scalar fields.

\section{One loop free energy in $AdS_4$ Vasiliev's theory}

Let us first consider the bosonic type A Vasiliev's theory whose spectrum consists of massless higher spin fields of all integer spins $s=1,2,\ldots$, each occurring once, and one conformally coupled scalar field. When the scalar field is quantized with the $\Delta=1$ boundary condition\footnote{A conformally coupled scalar in 4d has $m^2=-2/\ell^2$ (we will set the AdS radius $\ell = 1$ in the following), which leads to dual conformal dimensions $\Delta=1,2$.} and all higher spin fields with ordinary boundary conditions $\Delta=s+1$, this model is conjecturally dual to the 3d theory of $N$ free complex scalars in the $U(N)$ singlet sector \cite{Klebanov:2002ja}. The bulk one loop partition function is
\es{Z-1loop-4d}{
Z_{\rm 1-loop} =\frac{1}{\left[{\rm \det}\left(-\nabla^2-2\right)\right]^{\frac{1}{2}}}
\prod_{s=1}^{\infty}\frac{\left[{\rm det}^{STT}_{s-1}\left(-\nabla^2+s^2-1\right)\right]^{\frac{1}{2}}}{\left[{\rm det}^{STT}_{s}\left(-\nabla^2+s(s-2)-2\right)\right]^{\frac{1}{2}}} \ .
}
The structure of the higher spin determinant arises from gauge fixing, and includes the contribution of the anticommuting spin $s-1$ ghosts \cite{Gibbons:1978ac, Gibbons:1978ji, Christensen:1979iy, Yasuda:1983hk,Gaberdiel:2010ar, Gaberdiel:2010xv, Gupta:2012he}. The label STT indicates that the determinants are evaluated on the space of symmetric traceless transverse tensors. These one loop determinants can be computed with the aid of the spectral zeta function \cite{Hawking:1976ja} which was derived for all integer spins and all dimensions by Camporesi and Higuchi \cite{Camporesi:1993mz,Camporesi:1994ga}. Given the differential operator $(-\nabla^2+\kappa^2)$ (with $\kappa$ a constant) acting on the space of STT spin $s$ fields, one can define an associated spectral zeta function as the Mellin transform of the corresponding heat kernel. Explicitly, in general boundary dimension $d$ one finds \cite{Camporesi:1993mz,Camporesi:1994ga}
\begin{eqnarray}
&&\zeta_{(\Delta,s)}(z)=\left({ \int \vol_{AdS_{d+1} } \over \int \vol_{S^{d}}} \right) {2^{d-1} \over \pi} g(s) \int_0^\infty d u \, { \mu_s(u) \over \left[ u^2 + \left( \Delta - {d \over 2} \right)^2 \right]^z } \cr
&&\left(\Delta-\frac{d}{2}\right)^2=\kappa^2+s+\frac{d^2}{4}\,.
\label{zeta-d}
\end{eqnarray}
Here $\vol_{AdS_{d+1}}$ is the (regularized) volume of Euclidean AdS, $g(s)$ is the number of degrees of freedom of a STT spin $s$ field in $d+1$ dimensions, and $\mu_s(u)$ is the so-called spectral density. In the present case of $d=3$, we have
\begin{eqnarray}
&&\vol_{AdS_{4}} = \frac{4}{3}\pi^2\,,\qquad \vol_{S^3}=2\pi^2 \cr
&&\mu_s(u) =\frac{\pi u}{16}\left[u^2+\left(s+\frac{1}{2}\right)^2\right]\tanh\pi u\,,\qquad g(s)=2s+1\,.
\end{eqnarray}
Given the spectral zeta function (\ref{zeta-d}), the contribution to the one loop free energy $F^{(1)}=-\log Z_{\rm 1-loop}$  of the spin $s$ field with kinetic operator $(-\nabla^2+\kappa^2)$ is obtained as \cite{Hawking:1976ja}
\begin{equation}
F^{(1)}_{(\Delta,s)} = -\frac{1}{2}\zeta'_{(\Delta,s)}(0)-\frac{1}{2}\zeta_{(\Delta,s)}(0) \log\left(\ell^2\Lambda^2\right)\,,
\label{F-one-loop}
\end{equation}
where $\Lambda$ is a renormalization mass scale. The logarithmic term proportional to $\zeta_{(\Delta,s)}(0)$ arises in even dimensional spacetimes and is related to the conformal anomaly. We now show that this logarithmic divergence vanishes in Vasiliev's theory when we use the $\zeta$ function regularization to sum over all spins.
The value of the spectral zeta function at $z=0$ can be computed by writing
\begin{equation}
\tanh\pi u=1-\frac{2}{e^{2\pi u}+1}
\end{equation}
and evaluating the resulting integral by analytic continuation in $z$. This leads to \cite{Camporesi:1993mz}
\begin{equation}
\zeta_{(\Delta,s)}(0)=\frac{1}{24}(2s+1)\left[\nu^4-\left(s+\frac{1}{2}\right)^2\left(2\nu^2+\frac{1}{6}\right)
-\frac{7}{240}\right]\,,\qquad \nu\equiv \Delta-\frac{3}{2}\,.
\end{equation}
From (\ref{Z-1loop-4d}), we see that $\Delta=s+1$ for the physical transverse spin $s$ field, and $\Delta=s+2$ for the spin $s-1$ ghosts. Then the full logarithmic term of the one loop free energy is
\begin{eqnarray}
F^{(1)}\Big{|}_{\rm log-div} &=& -\frac{1}{2} \left(\zeta_{(1,0)}(0)+\sum_{s=1}^{\infty}\left(\zeta_{(s+1,s)}(0)-\zeta_{(s+2,s-1)}(0)\right)\right)\log\left(\ell^2\Lambda^2\right)\cr
&=& \left(\frac{1}{360}+\sum_{s=1}^{\infty} \left(\frac{1}{180}-\frac{s^2}{24}+\frac{5s^4}{24}\right)\right)\log\left(\ell^2\Lambda^2\right)
\label{log-div}
\end{eqnarray}
Performing the sum over spins with $\zeta$ function regularization, and using $\zeta(0)=-\frac{1}{2}$ and $\zeta(-2n)=0$ for $n>0$, we see that the coefficient of the logarithmic divergence indeed vanishes. The cancelation of the odd powers of $s$ arises after subtracting the ghost contributions. As a test of (\ref{log-div}), one can see that for $s=0$ and $s=1$ one recovers the correct conformal anomaly coefficients for a conformal scalar and Maxwell field. For $s=2$, (\ref{log-div}) gives $\frac{571}{180}\log\left(\ell^2\Lambda^2\right)$, which is indeed the correct coefficient of the logarithmic divergence in pure Einstein gravity in $AdS_4$ \cite{Christensen:1979iy} (see also \cite{Bastianelli:2013tsa}).\footnote{The one-loop effective action for higher spin fields in AdS was also recently studied in \cite{Bastianelli:2012bn} using a worldline approach. The result of \cite{Bastianelli:2012bn} for the logarithimic divergence of a spin $s$ field in $AdS_4$ differs from (\ref{log-div}), except at $s=0,1$, however the $\zeta$-regularized sum over spins still vanishes. It would be interesting to understand the origin of this difference.}

Having shown that the logarithmic piece vanishes, we now move to the computation of the finite contribution to $F^{(1)}$. Evaluating the derivative of the spectral zeta function at $z=0$, one obtains \cite{Camporesi:1993mz}
\begin{equation}
\zeta'_{(\Delta,s)}(0)=\frac{1}{3}(2s+1)\left[\frac{\nu^4}{8}+\frac{\nu^2}{48}+c_1
+\left(s+\frac{1}{2}\right)^2 c_2+\int_0^{\nu} dx\left[\left(s+\frac{1}{2}\right)^2 x-x^3\right] \psi(x+\frac{1}{2})\right]\,
\label{zprime-DS}
\end{equation}
where $\psi(x)=\frac{d}{dx}\log\Gamma(x)$ is the digamma function, and $c_1,c_2$ are constants given by the integrals
\begin{equation}
c_1 = \int_0^{\infty}du \frac{u^3\log u^2}{e^{2\pi u}+1}\,,\qquad
c_2 = \int_0^{\infty}du \frac{u \log u^2}{e^{2\pi u}+1}\,.
\end{equation}
Let us define
\begin{equation}
{\cal I}(\nu,s)=\frac{1}{3}(2s+1)\int_0^{\nu} dx\left[\left(s+\frac{1}{2}\right)^2 x-x^3\right] \psi(x+\frac{1}{2})\,.
\label{calI}
\end{equation}
Then, for the $\Delta=1$ scalar we have\footnote{While the derivation of (\ref{zprime-DS}) from (\ref{zeta-d}) strictly speaking assumes positive $\nu=\Delta-3/2$, i.e. the $\Delta_{+}=2$ boundary condition, a consistent prescription \cite{Giombi:2013yva} to obtain the correct value for the alternate boundary condition $\Delta_{-}=1$ is to analytically continue (\ref{zprime-DS}) to negative $\nu$. This procedure gives a result in agreement with the difference $F_{\Delta_{+}}-F_{\Delta_{-}}$ computed using different methods \cite{Gubser:2002zh,Diaz:2007an,Klebanov:2011gs}.}
\begin{equation}
\zeta'_{(1,0)}(0)=\frac{1}{3}\left(\frac{5}{384}+c_1+\frac{c_2}{4}\right)+{\cal I}\left(-\frac{1}{2},0\right)\,,
\label{zprime-scalar}
\end{equation}
while the massless higher spin fields contribute
\begin{eqnarray}
\label{zprime-HS}
\zeta'_{(s+1,s)}(0)-\zeta'_{(s+2,s-1)}(0)
&=&\frac{1}{3}\left[\frac{5}{192}+2c_1+c_2\left(6s^2+\frac{1}{2}\right)+\frac{s^2}{12}-3 \frac{s^4}{4}\right]\\
&+&{\cal I}\left(s-\frac{1}{2},s\right)-{\cal I}\left(s+\frac{1}{2},s-1\right)\nonumber\,.
\end{eqnarray}
From these expression we see that the first three terms in (\ref{zprime-scalar}) are precisely canceled by the first line of (\ref{zprime-HS}) when we sum over all spins using the $\zeta$ function regularization. We can therefore concentrate on the remaining more complicated contributions. For the scalar, we find
\begin{equation}
{\cal I}\left(-\frac{1}{2},0\right)=-\frac{1}{3}\int_{-1/2}^{0} dx\left(\frac{x}{4}-x^3\right) \psi(x+\frac{1}{2})=
\frac{11}{1152}-\frac{11\log 2}{2880}-\frac{\log A}{8}-\frac{\zeta(3)}{8\pi^2}+\frac{5\zeta'(-3)}{8},
\label{calI-scalar}
\end{equation}
where $A=e^{\frac{1}{12}-\zeta'(-1)}$ is the Glaisher-Kinkelin constant. To compute the higher-spin contributions, it is convenient to use the integral representation of the digamma function
\begin{equation}
\psi(y) = \int_0^{\infty} dt \left(\frac{e^{-t}}{t}-\frac{e^{-y t}}{1-e^{-t}}\right)\,.
\end{equation}
Plugging this into (\ref{calI}) one can first perform the $x$-integral, and then sum over all $s$ with $\zeta$ function regularization. This yields the result
\begin{eqnarray}
&&\sum_{s=1}^{\infty}\left[{\cal I}\left(s-\frac{1}{2},s\right)-{\cal I}\left(s+\frac{1}{2},s-1\right)\right]\cr
\label{big-t-int}
&&=\int_0^{\infty} dt\Bigg{[} \frac{191 e^{-t} +1349 e^{-2t}+1334 e^{-3t}+202 e^{-4t}-5 e^{-5t}+e^{-6t}}{192(1-e^{-t})^5 t} \\
&&+\frac{e^{-\frac{t}{2}}+18e^{-t}-e^{-\frac{3}{2}t}-2e^{-2t}}{12(1-e^{-t})^2 t^2}
-\frac{3e^{-t}+6 e^{-2t}-e^{-3t}}{{(1-e^{-t})^3 t^3}}
-2\frac{e^{-\frac{t}{2}}+3 e^{-t}-e^{-\frac{3}{2}t}+e^{-2t}}{(1-e^{-t})^2 t^4} \Bigg{]}\,.
\nonumber
\end{eqnarray}
This integral has only cubic and linear divergences near $t=0$, where the integrand behaves as $\frac{8}{3t^4}-\frac{1}{9t^2}+O(t^0)$. Note that, importantly, the logarithmic divergence is absent. We will compute this integral by analytic continuation, in the same spirit as the $\zeta$ function regularization. This can be done by using the integral representation of the Hurwitz-Lerch function
\begin{equation}
\Phi(z,s,v)=\frac{1}{\Gamma(s)}\int_0^{\infty} dt \frac{t^{s-1}e^{-v t}}{1-z e^{-t}}=\sum_{n=0}^{\infty} (n+v)^{-s} z^n\,.
\end{equation}
The integral (\ref{big-t-int}) can be expressed as a combination of $\Phi(z,s,v)$ and its $z$-derivatives evaluated at $z=1$. In turn, these can be related to the Hurwitz zeta function
\begin{equation}
\zeta(s,v)=\sum_{n=0}^{\infty} (n+v)^{-s}\,,
\end{equation}
which can be easily analytically continued to negative $s$ analogously to the ordinary Riemann zeta function. After a somewhat lengthy calculation, we find the result
\begin{eqnarray}
&&\sum_{s=1}^{\infty}\left[{\cal I}\left(s-\frac{1}{2},s\right)-{\cal I}\left(s+\frac{1}{2},s-1\right)\right]\cr
&&= -\frac{11}{1152}+\frac{11\log 2}{2880}+\frac{\log A}{8}
-\frac{5\zeta'(-3)}{8}-\frac{\zeta'(-2)}{2}\,.
\end{eqnarray}
Noting that $\zeta'(-2)=-\frac{\zeta(3)}{4\pi^2}$, we see that this precisely cancels the scalar contribution (\ref{calI-scalar})!\  As a further check, we have also evaluated the integral (\ref{big-t-int}) numerically after subtracting from the integrand the terms $\frac{8}{3t^4}-\frac{1}{9t^2}$ which give pure
power divergences; this procedure gives precise agreement with the analytic answer above. We conclude that the one loop vacuum energy in the bosonic type A Vasiliev theory with a $\Delta=1$ scalar precisely vanishes:
\begin{equation}
F^{(1)} = 0\,.
\end{equation}
This is consistent with the conjectured duality \cite{Klebanov:2002ja} with $N$ free complex scalars in the $U(N)$ singlet sector, and the simplest identification of the bulk coupling $G_N^{-1}=\gamma N$ with no order one shifts of $N$. The results of \cite{Diaz:2007an,Klebanov:2011gs,Giombi:2013yva}, or a direct evaluation of (\ref{calI}) with $\nu=1/2,s=0$, then imply that when the scalar is quantized with $\Delta=2$ boundary condition $F^{(1)} = - \frac {\zeta(3)} {8 \pi^2}$, consistent with the field theory containing
the double-trace interaction $\sim (\bar \phi^a \phi^a)^2$.
Of course, it remains to be shown that $G_N^{-1} F^{(0)} = N\left(\frac{\log 2}{4}-\frac{3\zeta(3)}{8\pi^2}\right)$.

We can perform an analogous calculation in the case of the ``minimal" type A Vasiliev's theory which contains one conformally coupled scalar and one field of each even spin $s=2,4,\ldots$. For the $\Delta=1$ boundary condition on the bulk scalar, this theory is conjecturally dual to $N$ free real scalars in the $O(N)$ singlet sector \cite{Klebanov:2002ja}. The computation proceeds along the same lines as described above. First one finds that, as in (\ref{log-div}), the logarithmic divergence cancels. This is because in $\zeta$ function regularization $\sum_{{\rm even}\,s}\left(\frac{1}{180}-\frac{s^2}{24}+\frac{5s^4}{24}\right)=-\frac{1}{360}$. The finite contribution to the one loop free energy is given by
\begin{equation}
F^{(1)}_{{\rm min}} = -\frac{1}{2}{\cal I}\left(-\frac{1}{2},0\right)-\frac{1}{2}\sum_{{\rm even}\,s\ge 2} \left[{\cal I}\left(s-\frac{1}{2},s\right)-{\cal I}\left(s+\frac{1}{2},s-1\right)\right]\,.
\end{equation}
The sum over spins can be evaluated as explained above, which leads to the integral
\begin{eqnarray}
&&\!\!\!\!\!\!\!\!\!\!\!\!\!\!\!\!
\sum_{{\rm even}\,s \ge 2} \left[{\cal I}\left(s-\frac{1}{2},s\right)-{\cal I}\left(s+\frac{1}{2},s-1\right)\right]\cr
\label{big-t-int-even}
&&\!\!\!\!\!\!\!\!\!\!\!\!\!\!\!\!=-\int_0^{\infty} dt\Bigg{[}2 \frac{e^{-\frac{t}{2}}+e^{-\frac{3}{2}t}+5e^{-2t}-e^{-\frac{5}{2}t}+2 e^{-3t}-e^{-\frac{7}{2}t}+e^{-4t}}{{(1-e^{-t})^2 (1+e^{-t})^2 t^4}}\\
&&\!\!\!\!\!\!\!\!\!\!\!\!\!\!\!\!+\frac{15 e^{-2t}+15 e^{-3t}+18 e^{-4t}+18 e^{-5t}-e^{-6t}-e^{-7t}}{(1-e^{-t})^3(1+e^{-t})^4 t^3}\cr
&&\!\!\!\!\!\!\!\!\!\!\!\!\!\!\!\!-\frac{e^{-\frac{t}{2}}+3 e^{-\frac{3}{2}t}-10 e^{-2t}+2 e^{-\frac{5}{2}t}+168 e^{-3t}-2 e^{-\frac{7}{2}t}-20 e^{-4t}-3 e^{-\frac{9}{2}t}-8 e^{-5t}-e^{-\frac{11}{2}t}-2 e^{-6t}}{12(1-e^{-t})^2 (1+e^{-t})^4 t^2}\cr
&&\!\!\!\!\!\!\!\!\!\!\!\!\!\!\!\!+\frac{e^{-t}-1921 e^{-2t}-1924 e^{-3t}-8444e^{-4t}-8442 e^{-5t}-1926e^{-6t}-1924 e^{-7t}+4 e^{-8t}+e^{-9t}-e^{-10t}}{192(1-e^{-t})^5 (1+e^{-t})^4 t} \Bigg{]}\,.
\nonumber
\end{eqnarray}
This can be computed similarly to (\ref{big-t-int}) by using the relation
\begin{equation}
\frac{1}{(1-e^{-t})^{n+1}(1+e^{-t})^{m+1}}=\frac{(-1)^{n}}{n!m!}\partial_{z_1}^n\partial_{z_2}^m\left[\frac{1}{z_1-z_2}
\left(\frac{1}{z_1-e^{-t}}-\frac{1}{z_2-e^{-t}}\right)\right]\Bigg{|}_{z_1=1,z_2=-1}
\end{equation}
and the integral representation of the Hurwitz-Lerch function. Now we also need its value
at $z=-1$, which is related to the
Hurwitz zeta function
through the identity $2^s \Phi(-1,s,v)=\zeta(s,\frac{v}{2})-\zeta(s,\frac{1+v}{2})$. The final result turns out to be
\begin{eqnarray}
&&\sum_{{\rm even}\,s\ge 2} \left[{\cal I}\left(s-\frac{1}{2},s\right)-{\cal I}\left(s+\frac{1}{2},s-1\right)\right]\cr
&&=-\frac{11}{1152}-\frac{709\log 2}{2880}+\frac{\log A}{8}+\frac{\zeta(3)}{2\pi^2}-\frac{5\zeta'(-3)}{8}\,.
\label{evensum}
\end{eqnarray}
To check the analytic continuation prescription, we have also evaluated the integral (\ref{big-t-int-even}) numerically after subtracting from the integrand the terms $\frac{4}{3t^4}-\frac{1}{18t^2}$ which give pure
power divergences; this gives precise agreement with the analytic answer above.
Combining (\ref{evensum}) with the scalar contribution (\ref{calI-scalar}), we get
\begin{equation}
F^{(1)}_{{\rm min}} = \frac{\log 2}{8}-\frac{3\zeta(3)}{16\pi^2}\,.
\end{equation}
Thus, we find that the one loop vacuum energy does not vanish in the minimal type A theory. However, remarkably, it is precisely equal to the value of the 3-sphere free energy $F$ for a real conformally coupled scalar field!\ This result is therefore consistent with the duality with $N$ free real scalars in the $O(N)$ singlet sector, provided the identification between the bulk coupling $G_N$ and $N$ involves a shift $N\rightarrow N-1$, namely
\begin{equation}
\frac{1}{G_N}F^{(0)}_{\rm min} = (N-1)\left(\frac{\log 2}{8}-\frac{3\zeta(3)}{16\pi^2}\right)\,.
\end{equation}
Interestingly, a similar shift $N\rightarrow N-1$ appears in the dictionary relating the `t Hooft coupling to bulk parameters in the duality between pure $SO(N)$ Chern-Simons theory and the topological string \cite{Sinha:2000ap}. It implies that the $SO(N)$ Chern-Simons theory has a natural large $N$ expansion in powers of $N-1$. Since a complete bulk dual to the singlet sector of the vector model would require coupling Vasiliev's theory to a topological sector describing the pure Chern-Simons dynamics, it is likely that the result we find is connected to that found in \cite{Sinha:2000ap}.

The above considerations imply a special role of the CFT with $N=1$, i.e. a single free real scalar field.\footnote{One could speculate that in the dual $AdS_4$ description this field arises as a singleton.
Perhaps there is a relation with the ideas in \cite{Leigh:2012mz,Gelfond:2013xt}.} The dual higher spin description of this theory involves infinite $G_N$, so that there is no classical contribution to $F$; nor is there any Chern-Simons contribution because $SO(1)$ is trivial.
Remarkably, the sum over one loop vacuum energies of even spin fields in Euclidean $AdS_4$ reproduces the exact value of $F$ in the free scalar field theory, and there should not be any
further corrections. Thus, the $N=1$ case of the duality proposed in \cite{Klebanov:2002ja} may be ``topological" and warrants special investigation.

\subsection{$USp(N)$ invariant 3-d scalar theory and its Vasiliev dual}

In this section we make a duality conjecture for the $husp(2;0|4)$ Vasiliev theory in $AdS_4$ and carry out its one loop test.
In general, the $husp(n;m|4)$ theory \cite{Vasiliev:1999ba} contains $\frac{n(n+1)+ m(m+1)}{2}$ massless gauge fields of odd spin,
$\frac{n(n-1)+ m(m-1)}{2}$ fields of even spin, and $nm$ fields of half-integral spin (here both $n$ and $m$ must be even).
For $n=m$ the theory is supersymmetric, while the minimal bosonic theory appears for $n=2$, $m=0$.
The spectrum of the $husp(2;0|4)$ theory in $AdS_4$ consists of one massless field of each even spin and three massless fields of each odd spin. The latter transform in the adjoint of $USp(2)=SU(2)$. In particular, the bulk theory includes a non-abelian spin 1 field with gauge group $SU(2)$.

Let us make a conjecture for the 3-d CFT dual to this theory. It is the $USp(N)$ singlet sector of the theory of $N$ massless complex scalar fields in
3-d, where $N$ is even.\footnote{In our notation the group $USp(N)$ is the subset of the $N\times N$ unitary matrices which satisfy $U^{\rm T} Q U=Q$, where $Q=\left( \begin{array}{cc}
0 & {\bf 1}_{\frac{N}{2}}  \\
-{\bf 1}_{\frac{N}{2}}  & 0  \end{array} \right) $, and $N$ is necessarily even.}
To make such a projection we couple the $N$ complex scalars, $\phi^a$, $a=1,\ldots N$,
to the $USp(N)$ Chern-Simons theory of level $k$ and send $k\rightarrow \infty$.
The $USp(N)$ invariant operators in this theory come in two varieties.
The first type involves indices contracted with the $N\times N$
identity matrix $\delta_{ab}$. For spin $0$ this is the scalar operator $\delta_{ab} \phi^a \bar \phi^b$, for spin $1$ this is the current
$J_\mu= i\delta_{ab} \phi^a \stackrel{\leftrightarrow}\partial_\mu \bar \phi^b$, for spin $2$ this is the stress-energy tensor, etc. Clearly, there is one such operator  for each
integer spin. Since $USp(N)\subset U(N)$, these currents are $USp(N)$ invariant. Additionally, we can form $USp(N)$ invariant conserved currents where the indices are contracted with the antisymmetric matrix $Q_{ab}$,
such as $J^Q_\mu= i Q_{ab} \phi^a \partial_\mu \phi^b$ for spin 1. Such complex currents exist only for odd spin (for even spin they are total derivatives). They are invariant under an $USp(N)$ transformation $\phi \rightarrow U\phi$, because $U^{\rm T}QU=Q$.
Therefore, we get two additional real currents for each odd spin, so that in total there are 3 currents for each odd spin and 1 current for each even spin. The 3 odd spin currents combine into the adjoint of a global $SU(2)$ symmetry,\footnote{To see this, we may view the $N$ free complex fields as $2N$ real fields. Then the free lagrangian has a global $SO(2N)$ symmetry, and there are associated conserved currents in the adjoint of $SO(2N)$. When $N$ is even, $SO(2N)$ has a subgroup $USp(N)\times SU(2)$, and the adjoint of $SO(2N)$ decomposes as $\bf{N(2N-1)} = (\bf{1},\bf{3})+(\bf{N(N+1)/2},\bf{1})+(\bf{N(N-1)/2-1},\bf{3})$ in terms of $USp(N)\times SU(2)$ representations. This means that when we gauge the $USp(N)$ subgroup of $SO(2N)$ to impose the singlet constraint, we get 3 currents which are in the adjoint of a leftover global $SU(2)$.} which corresponds to the $SU(2)$ gauge group in the bulk. This spectrum of 3-d currents thus matches the spectrum of massless gauge fields in the $husp(2;0|4)$ theory in $AdS_4$. Such a 3-d CS-matter theory is dual to the Vasiliev theory with the $\Delta=1$ boundary condition
for the scalar field. If we add to the action the operator $(\delta_{ab} \phi^a \bar \phi^b)^2$ then the 3-d field theory flows to another CFT
dual to the $AdS_4$ theory with the $\Delta=2$ boundary condition on the scalar. Similarly, by modifying the boundary conditions on higher-spin fields
we can obtain 3-d CFT's where certain currents are gauged \cite{Giombi:2013yva}.

Let us calculate the one loop free energy in the bulk for the choice where the scalar bilinear has $\Delta=1$.
Since the logarithmic divergence cancels for the theory with all integer spins and for the theory with only even spins, it
again cancels. The finite part is
\begin{eqnarray}
F^{(1)}_{husp(2;0|4)} &=& -\frac{1}{2}{\cal I}\left(-\frac{1}{2},0\right)-\frac{1}{2}\sum_{{\rm even}\,s\ge 2} \left[{\cal I}\left(s-\frac{1}{2},s\right)-{\cal I}\left(s+\frac{1}{2},s-1\right)\right]\cr
&-& \frac{3}{2} \sum_{{\rm odd}\,s\ge 1} \left[{\cal I}\left(s-\frac{1}{2},s\right)-{\cal I}\left(s+\frac{1}{2},s-1\right)\right]= -\left(\frac{\log 2}{4}-\frac{3\zeta(3)}{8\pi^2}\right),
\end{eqnarray}
which is precisely minus the value of the 3-sphere free energy $F$ of a complex scalar field. This is consistent with our conjecture, provided there is a shift $N\rightarrow N+1$ in the bulk coupling constant, so that
\begin{eqnarray}
\frac{1}{G_N} F^{(0)}_{husp(2;0|4)}+F^{(1)}_{husp(2;0|4)}&=&(N+1)\left(\frac{\log 2}{4}-\frac{3\zeta(3)}{8\pi^2}\right)-\left(\frac{\log 2}{4}-\frac{3\zeta(3)}{8\pi^2}\right)\cr
&=&N\left(\frac{\log 2}{4}-\frac{3\zeta(3)}{8\pi^2}\right)\,.
\end{eqnarray}
A similar shift $N\rightarrow N+1$ indeed also appears in the duality between pure $USp(N)$ Chern-Simons theory and the topological string \cite{Sinha:2000ap}.

\section{Correction to central charge from one loop determinants in $AdS_3$}

In this section we carry out a one loop test of the Gaberdiel-Gopakumar duality conjecture \cite{Gaberdiel:2010ar},
which relates the large $N$ limit of $W_N$ minimal models, i.e. 2-d
$\frac{SU(N)_k\otimes SU(N)_1}{SU(N)_{k+1}}$ coset CFTs, to higher-spin theory in $AdS_3$ \cite{Prokushkin:1998bq}. The central charge of these CFTs has
the large $N$ expansion
\es{c-WN}{
c(N,k)=(N-1)\left(1-\frac{N(N+1)}{(N+k)(N+k+1)}\right)=N(1-\lambda^2)\mp \lambda^3-1+{\cal O}(\frac{1}{N})\ ,
}
where the plus sign corresponds to the coupling identification $\lambda = \frac{N}{N+k}$, while the minus sign
corresponds to the identification $\lambda = \frac{N}{N+k+1}$. We will be able to match the ${\cal O}(N^0)$ term, but only if we use the CFT operator truncation
proposed by Chang and Yin \cite{Chang:2011mz}, and also by Gaberdiel and Gopakumar \cite{Gaberdiel:2012ku}. In other words,
we will consider Vasiliev's higher spin theory in $AdS_3$ whose spectrum includes massless higher spin fields with spins $s=2,3,\ldots$ and two real scalars with mass $m^2=\lambda^2-1$,
corresponding to the same dual conformal dimensions $\Delta_{\pm}=1 \pm \lambda$. We find that the plus sign in (\ref{c-WN}) corresponds to
$\Delta_-$, while the minus sign corresponds to $\Delta_+$.

We will compute the one loop correction to the free energy of the theory on the (Euclidean) $AdS_3$ vacuum solution
\es{AdS3}{
ds^2=d\rho^2+\sinh^2\rho\, d\Omega_2.}
This is related to the free energy of the boundary CFT on a round $S^2$, and from its logarithmic divergent piece we can extract the central charge of the model.
The one loop partition function in the bulk is given by the ratio of determinants
\es{Z-1loop}{
Z_{\rm 1-loop} =\left( \frac{1}{\sqrt{{\rm \det}\left(-\nabla^2+\lambda^2-1\right)}}\right)^2
\prod_{s=2}^{\infty}\frac{\left[{\rm det}^{STT}_{s-1}\left(-\nabla^2+s(s-1)\right)\right]^{\frac{1}{2}}}{\left[{\rm det}^{STT}_{s}\left(-\nabla^2+s(s-3)\right)\right]^{\frac{1}{2}}},
}
or, with $F_{\rm 1-loop}=-\log Z_{\rm 1-loop}$,
\es{F}{
&F_{\rm 1-loop} = 2\times \frac{1}{2}\log {\rm \det} \left(-\nabla^2+\lambda^2-1\right)\\
&+\frac{1}{2}\sum_{s=2}^{\infty}\left(\log {\rm \det}^{STT}_{s}\left(-\nabla^2+s(s-3)\right)-\log {\rm \det}^{STT}_{s-1}\left(-\nabla^2+s(s-1)\right)\right)
}
In each of the scalar factors we have a choice of $\Delta_+$ or $\Delta_{-}$ boundary condition. The structure of the higher spin terms arises from gauge fixing and includes the contribution of the spin $s-1$ ghosts.

The relevant one loop determinants can be computed as in the previous section with the aid of the spectral zeta function (\ref{zeta-d}), where we should now set $d=2$.
In this case, we have $g(s)=1$ for $s=0$ and $g(s)=2$ for $s\ge 1$ and the volume factors are given by
\es{volumes}{\vol_{\HH^{3}}=-2\pi \log R\,,\qquad \vol_{S^{2}}=4\pi\,,}
where $R$ is the radius of the $S^2$ located at a large cutoff $\rho=\rho_c$.
In $d=2$, the spectral density is particularly simple \cite{Camporesi:1994ga}
\es{mu}{
\mu_s(u)=u^2+s^2\,.
}
The determinant of the operator $(-\nabla^2+\kappa^2)$ (with $\kappa$ a constant) acting on the space of symmetric traceless transverse spin $s$ fields is obtained as \footnote{In odd-dimensional space-time the logarithmic divergence proportional to $\zeta_{(\Delta,s)}(0)$ is absent.}
\begin{eqnarray}
\log {\rm det}^{STT}_s \left(-\nabla^2+\kappa^2\right)=-\zeta_{(\Delta,s)}'(0)\,,\qquad
\left(\Delta-1\right)^2=\kappa^2+s+1\,.
\end{eqnarray}
From the structure of the kinetic operators in (\ref{F}) we see that $\Delta=s$ for the spin $s$ fields and $\Delta=s+1$ for the spin $s-1$ ghosts (we choose the standard boundary conditions for all the higher spin fields), and $\Delta_{\pm}$ for the scalar. Evaluating the integral (\ref{zeta-d}), we get
\es{zprime}{
-\zeta_{(\Delta,s)}'(0)=\frac{1}{3}g(s)(\Delta-1)((\Delta-1)^2 - 3s^2)\log R\,.
}
So we find for $s\ge 2$
\es{logdet}{
\log {\rm \det}^{STT}_{s}\left(-\nabla^2+s(s-3)\right)-\log {\rm \det}^{STT}_{s-1}\left(-\nabla^2+s(s-1)\right)=-\frac{2}{3}(1+6 s(s-1))\log R
}
and for the scalar
\es{logdetSc}{
\log {\rm \det} \left(-\nabla^2+\lambda^2-1\right) = \pm \frac{\lambda^3}{3}\log R\,,
}
where $\pm$ corresponds to $\Delta_{\pm}=1\pm\lambda$. The full one loop free energy is then
\begin{eqnarray}
F_{\rm 1-loop}&=& \frac{1}{2}\sum_{s=2}^{\infty}\left(-\frac{2}{3}(1+6 s(s-1))\log R\right)+\frac{1}{2}\left(\pm 1 \pm 1\right)\frac{\lambda^3}{3}\log R \cr
&=&\frac{1}{3}\log R +\frac{1}{6}\left(\pm 1 \pm 1\right)\lambda^3\log R
\end{eqnarray}
where we have used $\zeta$ function regularization to perform the sum over all spins.\footnote{The contributions of fields with $s=2, 3, \ldots$ are the same,
up to a factor of $2$, as those found in \cite{Giombi:2013yva} for the difference between standard and alternate boundary conditions on higher spin fields. This is because (\ref{zprime}) is odd under the exchange $\Delta \rightarrow 2-\Delta$, so that taking the difference between $\Delta_{+}^{(s)}$ and $\Delta_{-}^{(s)}$ boundary conditions is the same as doubling the $\Delta_{+}^{(s)}$ result.
Similarly, using $AdS_5$ the spin $s$ contribution to the anomaly $a$-coefficient is
$-\frac{1}{360} s^2 (1+s)^2 [3+ 14 s (1+s)]$, which is again proportional to the result in
\cite{Giombi:2013yva}. The $\zeta$ function regularized sum of the $s=1, 2, \ldots$ contributions vanishes, and a $\Delta=2$ scalar field does not contribute.
Thus, the $\CO(N^0)$ correction to $a$-anomaly vanishes for this spectrum in $AdS_5$, in agreement with the theory of $N$ free complex conformally coupled
scalar fields in 4-d. Furthermore, the $\zeta$-function regularized sum over even spins gives $1/90$ which is the $a$-coefficient for a real conformal scalar.
}
Adopting the $\Delta_\pm$ boundary condition for both of the scalar fields, this yields
\es{F-res}{
F_{\rm 1-loop}=-\frac{1}{3}\left(\mp \lambda^3-1\right)\log R\,.
}
From the relation $F=-\frac{c}{3}\log R$ (here $c$ is normalized so that $c=1$ for a real free boson) we see that this agrees with the ${\cal O}(N^0)$ term in the large $N$ expansion of the central charge of the $W_N$ minimal model, (\ref{c-WN}). If we instead took one of the scalars to have $\Delta_-$ and the other $\Delta_+$
boundary condition, then the $\lambda$ dependence would cancel in the bulk calculation, and there would be no agreement with the CFT central charge.

A consistency check of the sign in the $\lambda^3$ term and the choice of $\Delta_{\pm}$ boundary condition can be made using the $c$-theorem. We can flow from the $\Delta_{-}$ (UV) theory to the $\Delta_{+}$ (IR) theory by a double-trace deformation, under which the central charge changes by $\delta c =-2\lambda^3$ \cite{Gaberdiel:2010pz}, in accordance with $c_{UV} > c_{IR}$. In other words, the $\Delta_{-}$ theory has a higher central charge. The $\lambda^3$ term in the ${\cal O}(N^0)$ piece of the central charge (\ref{c-WN}) has positive (negative) sign, and it only comes from the bulk scalar contribution. This shows why we must choose the $\Delta_{-}$
($\Delta_+$) boundary condition in order to reproduce this sign.

\section*{Acknowledgments}
We thank M. Gaberdiel, R. Gopakumar, B. Safdi, M. Vasiliev and X. Yin for helpful comments. The work of IRK was supported in part by the US NSF under Grants No.~PHY-0756966 and PHY-1314198. The work of SG was supported in part by NSF grant PHY-1318681.
\bibliographystyle{ssg}
\bibliography{CGLP}

\begingroup\raggedright\begin{thebibliography}{10}

\bibitem{Fradkin:1987ks}
E.~Fradkin and M.~A. Vasiliev, ``{On the Gravitational Interaction of Massless
  Higher Spin Fields},'' {\em Phys.Lett.} {\bf B189} (1987) 89--95.

\bibitem{Vasiliev:1990en}
M.~A. Vasiliev, ``{Consistent equation for interacting gauge fields of all
  spins in (3+1)-dimensions},'' {\em Phys.Lett.} {\bf B243} (1990) 378--382.

\bibitem{Vasiliev:1992av}
M.~A. Vasiliev, ``{More on equations of motion for interacting massless fields
  of all spins in (3+1)-dimensions},'' {\em Phys. Lett.} {\bf B285} (1992)
  225--234.

\bibitem{Vasiliev:1995dn}
M.~A. Vasiliev, ``{Higher-spin gauge theories in four, three and two
  dimensions},'' {\em Int. J. Mod. Phys.} {\bf D5} (1996) 763--797,
  \href{http://xxx.lanl.gov/abs/hep-th/9611024}{{\tt hep-th/9611024}}.

\bibitem{Vasiliev:1999ba}
M.~A. Vasiliev, ``{Higher spin gauge theories: Star-product and AdS space},''
  \href{http://xxx.lanl.gov/abs/hep-th/9910096}{{\tt hep-th/9910096}}.

\bibitem{Maldacena:1997re}
J.~M. Maldacena, ``{The Large $N$ Limit of Superconformal Field Theories and
  Supergravity},'' {\em Adv. Theor. Math. Phys.} {\bf 2} (1998) 231--252,
  \href{http://xxx.lanl.gov/abs/hep-th/9711200}{{\tt hep-th/9711200}}.

\bibitem{Gubser:1998bc}
S.~S. Gubser, I.~R. Klebanov, and A.~M. Polyakov, ``{Gauge Theory Correlators
  from Non-Critical String Theory},'' {\em Phys. Lett.} {\bf B428} (1998)
  105--114, \href{http://xxx.lanl.gov/abs/hep-th/9802109}{{\tt
  hep-th/9802109}}.

\bibitem{Witten:1998qj}
E.~Witten, ``{Anti-de~Sitter Space and Holography},'' {\em Adv. Theor. Math.
  Phys.} {\bf 2} (1998) 253--291,
  \href{http://xxx.lanl.gov/abs/hep-th/9802150}{{\tt hep-th/9802150}}.

\bibitem{Klebanov:2002ja}
I.~R. Klebanov and A.~M. Polyakov, ``{AdS dual of the critical $O(N)$ vector
  model},'' {\em Phys. Lett.} {\bf B550} (2002) 213--219,
  \href{http://xxx.lanl.gov/abs/hep-th/0210114}{{\tt hep-th/0210114}}.

\bibitem{Sezgin:2003pt}
E.~Sezgin and P.~Sundell, ``{Holography in 4D (super) higher spin theories and
  a test via cubic scalar couplings},'' {\em JHEP} {\bf 07} (2005) 044,
  \href{http://xxx.lanl.gov/abs/hep-th/0305040}{{\tt hep-th/0305040}}.

\bibitem{Leigh:2003gk}
R.~G. Leigh and A.~C. Petkou, ``{Holography of the ${\cal N} = 1$ higher-spin
  theory on AdS$_4$},'' {\em JHEP} {\bf 06} (2003) 011,
  \href{http://xxx.lanl.gov/abs/hep-th/0304217}{{\tt hep-th/0304217}}.

\bibitem{Giombi:2009wh}
S.~Giombi and X.~Yin, ``{Higher Spin Gauge Theory and Holography: The
  Three-Point Functions},'' {\em JHEP} {\bf 1009} (2010) 115,
  \href{http://xxx.lanl.gov/abs/0912.3462}{{\tt 0912.3462}}.

\bibitem{Giombi:2010vg}
S.~Giombi and X.~Yin, ``{Higher Spins in AdS and Twistorial Holography},'' {\em
  JHEP} {\bf 1104} (2011) 086, \href{http://xxx.lanl.gov/abs/1004.3736}{{\tt
  1004.3736}}.

\bibitem{Giombi:2011ya}
S.~Giombi and X.~Yin, ``{On Higher Spin Gauge Theory and the Critical $O(N)$
  Model},'' \href{http://xxx.lanl.gov/abs/1105.4011}{{\tt 1105.4011}}.

\bibitem{Maldacena:2011jn}
J.~Maldacena and A.~Zhiboedov, ``{Constraining Conformal Field Theories with A
  Higher Spin Symmetry},'' \href{http://xxx.lanl.gov/abs/1112.1016}{{\tt
  1112.1016}}.

\bibitem{Maldacena:2012sf}
J.~Maldacena and A.~Zhiboedov, ``{Constraining conformal field theories with a
  slightly broken higher spin symmetry},'' {\em Class.Quant.Grav.} {\bf 30}
  (2013) 104003, \href{http://xxx.lanl.gov/abs/1204.3882}{{\tt 1204.3882}}.

\bibitem{Colombo:2012jx}
N.~Colombo and P.~Sundell, ``{Higher Spin Gravity Amplitudes From Zero-form
  Charges},'' \href{http://xxx.lanl.gov/abs/1208.3880}{{\tt 1208.3880}}.

\bibitem{Didenko:2012tv}
V.~Didenko and E.~Skvortsov, ``{Exact higher-spin symmetry in CFT: all
  correlators in unbroken Vasiliev theory},''
  \href{http://xxx.lanl.gov/abs/1210.7963}{{\tt 1210.7963}}.

\bibitem{Gelfond:2013xt}
O.~Gelfond and M.~Vasiliev, ``{Operator algebra of free conformal currents via
  twistors},'' \href{http://xxx.lanl.gov/abs/1301.3123}{{\tt 1301.3123}}.

\bibitem{Didenko:2013bj}
V.~Didenko, J.~Mei, and E.~Skvortsov, ``{Exact higher-spin symmetry in CFT:
  free fermion correlators from Vasiliev Theory},'' {\em Phys.Rev.} {\bf D88}
  (2013) 046011, \href{http://xxx.lanl.gov/abs/1301.4166}{{\tt 1301.4166}}.

\bibitem{Aharony:2011jz}
O.~Aharony, G.~Gur-Ari, and R.~Yacoby, ``{d=3 Bosonic Vector Models Coupled to
  Chern-Simons Gauge Theories},'' {\em JHEP} {\bf 1203} (2012) 037,
  \href{http://xxx.lanl.gov/abs/1110.4382}{{\tt 1110.4382}}.

\bibitem{Giombi:2011kc}
S.~Giombi, S.~Minwalla, S.~Prakash, S.~P. Trivedi, S.~R. Wadia, {\em et.~al.},
  ``{Chern-Simons Theory with Vector Fermion Matter},'' {\em Eur.Phys.J.} {\bf
  C72} (2012) 2112, \href{http://xxx.lanl.gov/abs/1110.4386}{{\tt 1110.4386}}.

\bibitem{Chang:2012kt}
C.-M. Chang, S.~Minwalla, T.~Sharma, and X.~Yin, ``{ABJ Triality: from Higher
  Spin Fields to Strings},'' \href{http://xxx.lanl.gov/abs/1207.4485}{{\tt
  1207.4485}}.

\bibitem{Giombi:2012ms}
S.~Giombi and X.~Yin, ``{The Higher Spin/Vector Model Duality},''
  \href{http://xxx.lanl.gov/abs/1208.4036}{{\tt 1208.4036}}.

\bibitem{Klebanov:2011gs}
I.~R. Klebanov, S.~S. Pufu, and B.~R. Safdi, ``{$F$-Theorem without
  Supersymmetry},'' {\em JHEP} {\bf 1110} (2011) 038,
  \href{http://xxx.lanl.gov/abs/1105.4598}{{\tt 1105.4598}}.

\bibitem{Boulanger:2011dd}
N.~Boulanger and P.~Sundell, ``{An action principle for Vasiliev's
  four-dimensional higher-spin gravity},'' {\em J.Phys.} {\bf A44} (2011)
  495402, \href{http://xxx.lanl.gov/abs/1102.2219}{{\tt 1102.2219}}.

\bibitem{Boulanger:2012bj}
N.~Boulanger, N.~Colombo, and P.~Sundell, ``{A minimal BV action for Vasiliev's
  four-dimensional higher spin gravity},'' {\em JHEP} {\bf 1210} (2012) 043,
  \href{http://xxx.lanl.gov/abs/1205.3339}{{\tt 1205.3339}}.

\bibitem{Doroud:2011xs}
N.~Doroud and L.~Smolin, ``{An Action for higher spin gauge theory in four
  dimensions},'' \href{http://xxx.lanl.gov/abs/1102.3297}{{\tt 1102.3297}}.

\bibitem{Vasiliev:1988sa}
M.~A. Vasiliev, ``{CONSISTENT EQUATIONS FOR INTERACTING MASSLESS FIELDS OF ALL
  SPINS IN THE FIRST ORDER IN CURVATURES},'' {\em Annals Phys.} {\bf 190}
  (1989) 59--106.

\bibitem{Gubser:2002zh}
S.~S. Gubser and I.~Mitra, ``{Double trace operators and one loop vacuum energy
  in AdS / CFT},'' {\em Phys.Rev.} {\bf D67} (2003) 064018,
  \href{http://xxx.lanl.gov/abs/hep-th/0210093}{{\tt hep-th/0210093}}.

\bibitem{Gubser:2002vv}
S.~S. Gubser and I.~R. Klebanov, ``{A Universal result on central charges in
  the presence of double trace deformations},'' {\em Nucl.Phys.} {\bf B656}
  (2003) 23--36, \href{http://xxx.lanl.gov/abs/hep-th/0212138}{{\tt
  hep-th/0212138}}.

\bibitem{Hartman:2006dy}
T.~Hartman and L.~Rastelli, ``{Double-trace deformations, mixed boundary
  conditions and functional determinants in AdS/CFT},'' {\em JHEP} {\bf 0801}
  (2008) 019, \href{http://xxx.lanl.gov/abs/hep-th/0602106}{{\tt
  hep-th/0602106}}.

\bibitem{Diaz:2007an}
D.~E. Diaz and H.~Dorn, ``{Partition functions and double-trace deformations in
  AdS/CFT},'' {\em JHEP} {\bf 0705} (2007) 046,
  \href{http://xxx.lanl.gov/abs/hep-th/0702163}{{\tt hep-th/0702163}}.

\bibitem{Giombi:2013yva}
S.~Giombi, I.~R. Klebanov, S.~S. Pufu, B.~R. Safdi, and G.~Tarnopolsky, ``{AdS
  Description of Induced Higher-Spin Gauge Theory},''
  \href{http://xxx.lanl.gov/abs/1306.5242}{{\tt 1306.5242}}.

\bibitem{Sinha:2000ap}
S.~Sinha and C.~Vafa, ``{SO and Sp Chern-Simons at large N},''
  \href{http://xxx.lanl.gov/abs/hep-th/0012136}{{\tt hep-th/0012136}}.

\bibitem{Banerjee:2012gh}
S.~Banerjee, S.~Hellerman, J.~Maltz, and S.~H. Shenker, ``{Light States in
  Chern-Simons Theory Coupled to Fundamental Matter},'' {\em JHEP} {\bf 1303}
  (2013) 097, \href{http://xxx.lanl.gov/abs/1207.4195}{{\tt 1207.4195}}.

\bibitem{JHV}
M.~Aganagic, D.~Jafferis, S.~Hellerman, and C.~Vafa, ``Work in progress,''.

\bibitem{Gopakumar:1998ki}
R.~Gopakumar and C.~Vafa, ``{On the gauge theory / geometry correspondence},''
  {\em Adv.Theor.Math.Phys.} {\bf 3} (1999) 1415--1443,
  \href{http://xxx.lanl.gov/abs/hep-th/9811131}{{\tt hep-th/9811131}}.

\bibitem{Aharony:2012nh}
O.~Aharony, G.~Gur-Ari, and R.~Yacoby, ``{Correlation Functions of Large N
  Chern-Simons-Matter Theories and Bosonization in Three Dimensions},''
  \href{http://xxx.lanl.gov/abs/1207.4593}{{\tt 1207.4593}}.

\bibitem{GurAri:2012is}
G.~Gur-Ari and R.~Yacoby, ``{Correlators of Large N Fermionic Chern-Simons
  Vector Models},'' {\em JHEP} {\bf 1302} (2013) 150,
  \href{http://xxx.lanl.gov/abs/1211.1866}{{\tt 1211.1866}}.

\bibitem{Niemi:1983rq}
A.~Niemi and G.~Semenoff, ``{Axial Anomaly Induced Fermion Fractionization and
  Effective Gauge Theory Actions in Odd Dimensional Space-Times},'' {\em
  Phys.Rev.Lett.} {\bf 51} (1983) 2077.

\bibitem{Redlich:1983kn}
A.~Redlich, ``{Gauge Noninvariance and Parity Violation of Three-Dimensional
  Fermions},'' {\em Phys.Rev.Lett.} {\bf 52} (1984) 18.

\bibitem{Gaberdiel:2010pz}
M.~R. Gaberdiel and R.~Gopakumar, ``{An $AdS_3$ Dual for Minimal Model CFTs},''
  {\em Phys.Rev.} {\bf D83} (2011) 066007,
  \href{http://xxx.lanl.gov/abs/1011.2986}{{\tt 1011.2986}}.

\bibitem{Gaberdiel:2012uj}
M.~R. Gaberdiel and R.~Gopakumar, ``{Minimal Model Holography},''
  \href{http://xxx.lanl.gov/abs/1207.6697}{{\tt 1207.6697}}.

\bibitem{Prokushkin:1998bq}
S.~Prokushkin and M.~A. Vasiliev, ``{Higher spin gauge interactions for massive
  matter fields in 3-D AdS space-time},'' {\em Nucl.Phys.} {\bf B545} (1999)
  385, \href{http://xxx.lanl.gov/abs/hep-th/9806236}{{\tt hep-th/9806236}}.

\bibitem{Chang:2011mz}
C.-M. Chang and X.~Yin, ``{Higher Spin Gravity with Matter in $AdS_3$ and Its
  CFT Dual},'' {\em JHEP} {\bf 1210} (2012) 024,
  \href{http://xxx.lanl.gov/abs/1106.2580}{{\tt 1106.2580}}.

\bibitem{Gaberdiel:2012ku}
M.~R. Gaberdiel and R.~Gopakumar, ``{Triality in Minimal Model Holography},''
  {\em JHEP} {\bf 1207} (2012) 127,
  \href{http://xxx.lanl.gov/abs/1205.2472}{{\tt 1205.2472}}.

\bibitem{Gibbons:1978ac}
G.~Gibbons, S.~Hawking, and M.~Perry, ``{Path Integrals and the Indefiniteness
  of the Gravitational Action},'' {\em Nucl.Phys.} {\bf B138} (1978) 141.

\bibitem{Gibbons:1978ji}
G.~Gibbons and M.~Perry, ``{Quantizing Gravitational Instantons},'' {\em
  Nucl.Phys.} {\bf B146} (1978) 90.

\bibitem{Christensen:1979iy}
S.~Christensen and M.~Duff, ``{Quantizing Gravity with a Cosmological
  Constant},'' {\em Nucl.Phys.} {\bf B170} (1980) 480.

\bibitem{Yasuda:1983hk}
O.~Yasuda, ``{On the one loop effective potential in quantum gravity},'' {\em
  Phys.Lett.} {\bf B137} (1984) 52.

\bibitem{Gaberdiel:2010ar}
M.~R. Gaberdiel, R.~Gopakumar, and A.~Saha, ``{Quantum $W$-symmetry in
  AdS$_3$},'' {\em JHEP} {\bf 1102} (2011) 004,
  \href{http://xxx.lanl.gov/abs/1009.6087}{{\tt 1009.6087}}.

\bibitem{Gaberdiel:2010xv}
M.~R. Gaberdiel, D.~Grumiller, and D.~Vassilevich, ``{Graviton 1-loop partition
  function for 3-dimensional massive gravity},'' {\em JHEP} {\bf 1011} (2010)
  094, \href{http://xxx.lanl.gov/abs/1007.5189}{{\tt 1007.5189}}.

\bibitem{Gupta:2012he}
R.~K. Gupta and S.~Lal, ``{Partition Functions for Higher-Spin theories in
  AdS},'' {\em JHEP} {\bf 1207} (2012) 071,
  \href{http://xxx.lanl.gov/abs/1205.1130}{{\tt 1205.1130}}.

\bibitem{Hawking:1976ja}
S.~Hawking, ``{Zeta Function Regularization of Path Integrals in Curved
  Space-Time},'' {\em Commun.Math.Phys.} {\bf 55} (1977) 133.

\bibitem{Camporesi:1993mz}
R.~Camporesi and A.~Higuchi, ``{Arbitrary spin effective potentials in anti-de
  Sitter space-time},'' {\em Phys.Rev.} {\bf D47} (1993) 3339--3344.

\bibitem{Camporesi:1994ga}
R.~Camporesi and A.~Higuchi, ``{Spectral functions and zeta functions in
  hyperbolic spaces},'' {\em J.Math.Phys.} {\bf 35} (1994) 4217--4246.

\bibitem{Bastianelli:2013tsa}
F.~Bastianelli and R.~Bonezzi, ``{One-loop quantum gravity from a worldline
  viewpoint},'' {\em JHEP} {\bf 1307} (2013) 016,
  \href{http://xxx.lanl.gov/abs/1304.7135}{{\tt 1304.7135}}.

\bibitem{Bastianelli:2012bn}
F.~Bastianelli, R.~Bonezzi, O.~Corradini, and E.~Latini, ``{Effective action
  for higher spin fields on (A)dS backgrounds},'' {\em JHEP} {\bf 1212} (2012)
  113, \href{http://xxx.lanl.gov/abs/1210.4649}{{\tt 1210.4649}}.

\bibitem{Leigh:2012mz}
R.~G. Leigh and A.~C. Petkou, ``{Singleton deformation of higher-spin theory
  and the phase structure of the three-dimensional O(N) vector model},''
  \href{http://xxx.lanl.gov/abs/1212.4421}{{\tt 1212.4421}}.

\end{thebibliography}\endgroup

\end{document}